\documentclass{JASSS}
	
\title{BotSim: Mitigating The Formation Of Conspiratorial Societies with Useful Bots}

\reviewcopy{false} 

\author[1]{Lynnette Hui Xian Ng}
\author[2]{Kathleen M. Carley}
\affil[1,2]{ Center for Computational Analysis of Social and Organizational Systems (CASOS), Software and Societal Systems, Carnegie Mellon University, 5000 Forbes Ave, Pittsburgh, PA 15213}

\email{lynnetteng@cmu.edu}

\usepackage{natbib}
	\setcitestyle{authoryear,round,aysep={}}
\usepackage{hyperref}

\begin{document}
\maketitle 



\begin{abstract}
Societies can become a conspiratorial society where there is a majority of humans that believe, and therefore spread, conspiracy theories. Artificial intelligence gave rise to social media bots that can spread conspiracies in an automated fashion. Currently, organizations combat the spread of conspiracies through manual fact-checking processes and the dissemination of counter-narratives. However, the effects of harnessing the same automation to create useful bots are not well explored. To address this, we create BotSim, an Agent-Based Model of a society in which useful bots are introduced into a small world network. These useful bots are: Info-Correction Bots, which correct bad information into good, and Good Bots, which put out good messaging. The simulated agents interact through generating, consuming and propagating information.
Our results show that, left unchecked, Bad Bots can create a conspiratorial society, and this can be mitigated by either Info-Correction Bots or Good Bots; however, Good Bots are more efficient and sustainable than Info-Correction Bots . Proactive good messaging is more resource-effective than reactive information correction. 
With our observations, we expand the concept of bots as a malicious social media agent towards automated social media agent that can be used for both good and bad purposes. These results have implications for designing communication strategies to maintain a healthy social cyber ecosystem.
\end{abstract}


\parano{}




\section{Introduction}
\label{sec:intro}
Social media bots, or bots for short, are automated users that can distribute content~\citep{ng2025social}. Studies on bots typically focus on the negative societal effects of bots, such as their spread of disinformation, interference in stock markets and political events, and manipulation of public opinion \citep{aldayel2022characterizing,benigni2018bot}.
A much overlooked area of bots is how their automated nature can be harnessed for social good. A small handful of studies documented useful bots in the crisis informatics realm, for situational awareness and organizing aid \citep{hofeditz2019meaningful}. But can bots also be harnessed for preventing the manipulation of public opinion and the spread of disinformation?

In this paper, we use an agent-based simulation to investigate the effect of useful bots in combating malicious bots and preventing the formation of a conspiratorial society. A conspiratorial society is a society where there is a large number of individuals who believe, and therefore share conspiratorial narratives \citep{featherstone2000obscure,dyrendal2014hidden}. Conspiracies can be damaging to society because the extreme belief in conspiracies can result in offline consequences. In 2016, the Pizzagate conspiracy  rumored that there was a child pedophilia ring that used a pizza store as the front, resulting in a vigilante shooting incident \citep{bleakley2023panic}. In 2020, the coronavirus vaccine conspiracy theory stated that vaccines are a government tactic to spy on the population, or that vaccines have microchip implants \citep{phillips2022hoaxes}, creating barriers that reduced the ability of governments to curb the pandemic \citep{romer2020conspiracy}.

While conspiracy theories share some characteristics with broader categories of misinformation and fake news, they exhibit distinct features that warrant specialized studies \citep{douglas2018conspiracy,douglas2019understanding}. First, conspiracy theories form interconnected belief systems where believers resist contradictory evidence through a cognitive bias process called motivated reasoning \citep{leman2013beliefs}. Second, conspiracy theories exhibit unique social media propagation patterns, demonstrating sustained circulation and evolution over time \citep{dow2021covid}. Third, conspiracy theories are more likely to inspire offline behavioral consequences \citep{rottweiler2022conspiracy,jolley2014effects}, from vaccine hesitancy to vigilante actions, making their study particularly crucial for the field of social cyber security \citep{carley2020social}, which aims to build a resilient social cyber infrastructure. This work focuses on conspiracy theories rather than general misinformation, investigating the mechanics of the formation of a conspiratorial society, which is a society where the conspiracy theories have become beliefs normalized across the population.

To examine the effects of useful bots on a society, we create BotSim, an agent-based model, that incorporates the mechanics of Bad Bots that spread conspiracies, and useful bots that perform information correction (Info-Correction Bots) and good messaging (Good Bots) tasks. Our results show that if Bad Bots are left unchecked, a conspiratorial society will form. The introduction of useful bots prevents a conspiratorial society, although the society will reach a state where Bad Humans are the majority (i.e., Human agents that spread conspiracies). Importantly, a smaller proportion (50\%) of Good Bots compared to Info-Correction Bots is required to prevent the formation of a conspiracy society, and the employment of sufficient Good Bots (Bad:Good Bots ratio $\geq$1:1.6) prevents Bad Humans from reaching the majority.
These experimental results offer insights into the design and deployment of communication strategies, in particular good messaging, to slow the rate of formation of conspiracy societies on key issues.


\section{Related Work}
Conspiracy theories thrive in environments of uncertainty and complex societal challenges \citep{douglas2019understanding}. Events like an unprecedented health pandemic (i.e., the 2020 coronavirus pandemic) or unexplained phenomena (i.e., the 2024 Dubai rainfall and flood) are prime ground for conspiracy theories. Conspiracy theories arise as explanations for people to make sense of the situation. Conspiracy beliefs are difficult to correct because they promote interconnected belief systems, where rejecting one aspect of the theory threatens a person's entire worldview \citep{goertzel1994belief}. 
Social networks facilitate the formation of conspiratorial societies because individuals tend to cluster with others who have similar beliefs. This homophilic behavior results in cohesive echo chambers that amplify the spread of conspiracy narratives \citep{bessi2015science}. In fact, the algorithmic design of social platforms reinforces homophily by exposing users to content that they prefer, accelerating the formation of conspiratorial communities. 

Empirical studies show that the underlying network topology of conspiracy diffusion can resemble a small-world structure, where there are tightly clustered communities bridged by short path lengths \citep{ch2015local,huang2011preventing}. This structure allows a small number of highly active accounts to inject information across otherwise separate clusters, accelerating cross-community spread and the network's exposure to conspiracy theories.

In social networks, automated bot users also contribute to the spread of information. Studies of posts on X consistently show that bots produce posts at a scale almost twice that of humans in major events. Bot accounts posted 2.6 to 3 times more messages than human accounts during the 2020 US Presidential Elections \citep{luceri2020down}, and 2 times more messages during the 2020 Coronavirus pandemic \citep{ng2025social}. Although most studies model bots as harmful agents that spread conspiracy theories \citep{jin2025prevention,ross2019social}, this study expands the concept of the automated user to harness its automation as part of the intervention of conspiratorial spread.

Current methods of combating conspiracy theories focus primarily on fact-checking and debunking. The International Fact-Checking Network coordinates worldwide efforts to combat misinformation across social media platforms, but the fact-checking and debunking are heavily reliant on volunteer efforts.
To speed up the fact-checking process, computational approaches for identifying and countering conspiracy content have emerged. Several automated systems use linguistic markers and narrative structures to identify conspiratorial content \citep{memon2020characterizing,choudhary2021linguistic}. Automated fact-checking methods with machine learning or graph approaches show promise with scaling detection efforts, but is often resource- and data-hungry \citep{shu2017fake,medeiros2020fake}.

At the same time, qualitative asymmetries in human information sharing further complicate interventions. Empirical research shows that individuals with stronger conspiratorial tendencies exhibit more uniform information sharing, while those with weaker leanings share more diverse content\citep{pennycook2021psychology}. Therefore, simulation studies are useful to model and predict the impact of the complex dynamics of information diffusion.

To make interventions more difficult, online social networks can sometimes enable conspiracy theories to spread faster than fact-checking mechanisms can respond \citep{vosoughi2018spread}, because these theories are memorable and can incite huge engagement. Institutional fact-checking efforts can fail to reach susceptible communities \citep{jack2017lexicon}. Therefore, the deployment of useful bots to automate such interventions, can potentially provide a larger message reach through the power of social networks. Simulations are a useful tool to showcase the theoretical efficacy of automated interventions: \citet{kumar2014detecting} built an agent-based framework to counteract misinformation flows in social networks, and \citet{varol2020journalists} showed that strategically deployed counter-narrative bots can effectively reduce the spread of misinformation in simulated networks.
We build on these works to demonstrate the effectiveness of useful bots in combating conspiracy theories. 

Good messaging is a complementary approach to fact-checking. Fact-checking targets specific conspiratorial narratives, while good messaging promotes accurate information. Fact-checking is a reactive action, while good messaging is a proactive action. Good messaging builds on the inoculation theory, a psychological framework that exposes individuals to weakened forms of misinformation and the corresponding refutations to develop resistance against future, stronger persuasive attempts \citep{compton2021inoculation}. Preemptively exposing people to weakened forms of information as inoculation \citep{van2017inoculating} and media literacy interventions can help combat conspiracy thinking \citep{huang2024media}. However, such methods that focus primarily on belief correction at the individual level, and are difficult to scale up. Therefore, there is a need to harness automated tools to combat conspiracy theories in the online space as they arise.

Agent Based Modeling (ABM) has been used as a means to understand the dynamics of information propagation in online societies, demonstrating the formation of polarization and echo-chambers in social networks \citep{betts2022effect,keijzer2024polarization,lu2024agents}. Such models enable the creation of agent entities that each have unique characteristics and heuristics of interaction with other agents and the environment. By simulating successive interactions between agents across time, ABM allows for the observation of emergent behaviors that connect micro-level individual agent behavior to macro-level patterns \citep{betts2022effect}. This approach has proven particularly valuable for evaluating countermeasures because it enables controlled experimentation without the ethical concerns associated with introducing actual changes into real social networks \citep{murdock2025simulating}.

ABM has also been used to model intervention strategies against harmful information on social media platforms. Several studies have developed sophisticated frameworks that incorporate heterogeneous agents with varying susceptibility to misinformation and diverse intervention mechanisms \citep{carragher2023simulation,abouzeid2024towards}. Other similar studies model use opinion influence as a proxy for reinforcement learning in a social network structure \citep{jin2025synthetic}. These models demonstrate that successful interventions depend on the timing, target selection and the interactions between different types of countermeasures. Recent research has also explored the use of automated bot agents for interventions strategies such as fact-checking, content moderation, and counter-messaging \citep{cisneros2019spread}. With this backdrop, we employ the ABM technique to study the effects of useful bots in disrupting the activities of bad bots.

\section{Model}

\subsection{Purpose and Overview}

Our simulation uses an Agent-Based Model, named BotSim, set up in a Small World network to simulate the consumption and dissemination of information by a community of agents. The Small World network structure is commonly used to model social networks \citep{fan2012opinion}, because this structure (1) shows high local clustering, which reflects the tendency for people to form tightly connected social groups, and (2) has a small characteristic path length, mimicking the random and rapid information dissemination \citep{watts1998collective}.

The primary purpose of BotSim is to investigate the conditions under which useful bots can effectively counteract the formation of conspiratorial societies. These useful bots perform actions of information correction and positive messaging. This leads us to the main research question: Can automated useful bots prevent the formation of conspiratorial societies where Bad Bots are present? Empirically, we define a conspiracy society when all the Human Agents in the simulation have turned to Bad Humans. In our investigation of this question, we analyzed the optimal deployment ratios between Bad Bots and useful bots to maximize intervention effectiveness, and how the two types of useful bots (Info-Correction versus Good Bots) compare in their resource efficiency.

The BotSim model simulates a dynamic social media information ecosystem where agents continuously generate, consume, and propagate information through their ego network connections. In our virtual simulation experiments, we vary the proportions of four different agent types and measure the time in simulation ticks that is required for the convergence of the simulation towards a conspiracy society. The design choices made in the model are grounded in empirical literature on conspiracy belief, cognitive biases, and information propagation. Our ABM leverages the results from static analytical models (see Table~\ref{tab:stylized_facts}) to capture temporal dynamics of belief formation and feedback loops that can stabilize or destabilize the information environment.

\subsection{Types of Agents and Information}
There are two types of agents that interact in a network within BotSim: Human Agents and Bot Agents. For each type of agents, there are sub-types of agents.

First, we have Human Agents. Human Agents are either in a Good or Bad state. The state that the Human Agents are in depends on the amount of each type of Information they consume. The two types of Human agents are:
\begin{enumerate}
    \item \textbf{Good Humans} represent normal humans that generate and consume information in a social network.
    \item \textbf{Bad Humans} represent humans that have become believers of conspiracy theories, and therefore only propagate Bad Information.
\end{enumerate}

Next, we have Bot Agents, of which we have three types:
\begin{enumerate}
    \item \textbf{Bad Bots} are malicious bot agents that disseminate conspiracy theories. The number of Bad Bots is $\alpha_1 \times$ (total number of Humans).
    \item \textbf{Info-Correction Bots}, short for information-correction bots, that represent agents who fact-check false information and present corrected information. The number of Info-Correction Bots is  $\alpha_2 \times$ (number of Bad Bots)
    \item \textbf{Good Bots} represent agents that produce good messaging. The number of Good Bots is $\alpha_3 \times$ (number of Bad Bots).
\end{enumerate}
In this simulation, we assume that the Info-Correction and Good Bots are 100\% secure. That is these agent types are will always be reliable for their fact-checking and positive messaging activities respectively.
Visually, the agents are represented by the person stick figure. 

Agents consume and disseminate Information. There are two types of Information in our environment: 
\begin{enumerate}
    \item \textbf{Good Info}, which represents factual information. Good Info is visually represented by green translucent circles.
    \item \textbf{Bad Info}, which represents conspiracy theories. Bad Information is visually represented by red translucent circles on the person icons.
\end{enumerate}

\subsection{The model's main routine}

\begin{figure}[h]
    \centering
    \includegraphics[width=1.0\textwidth]{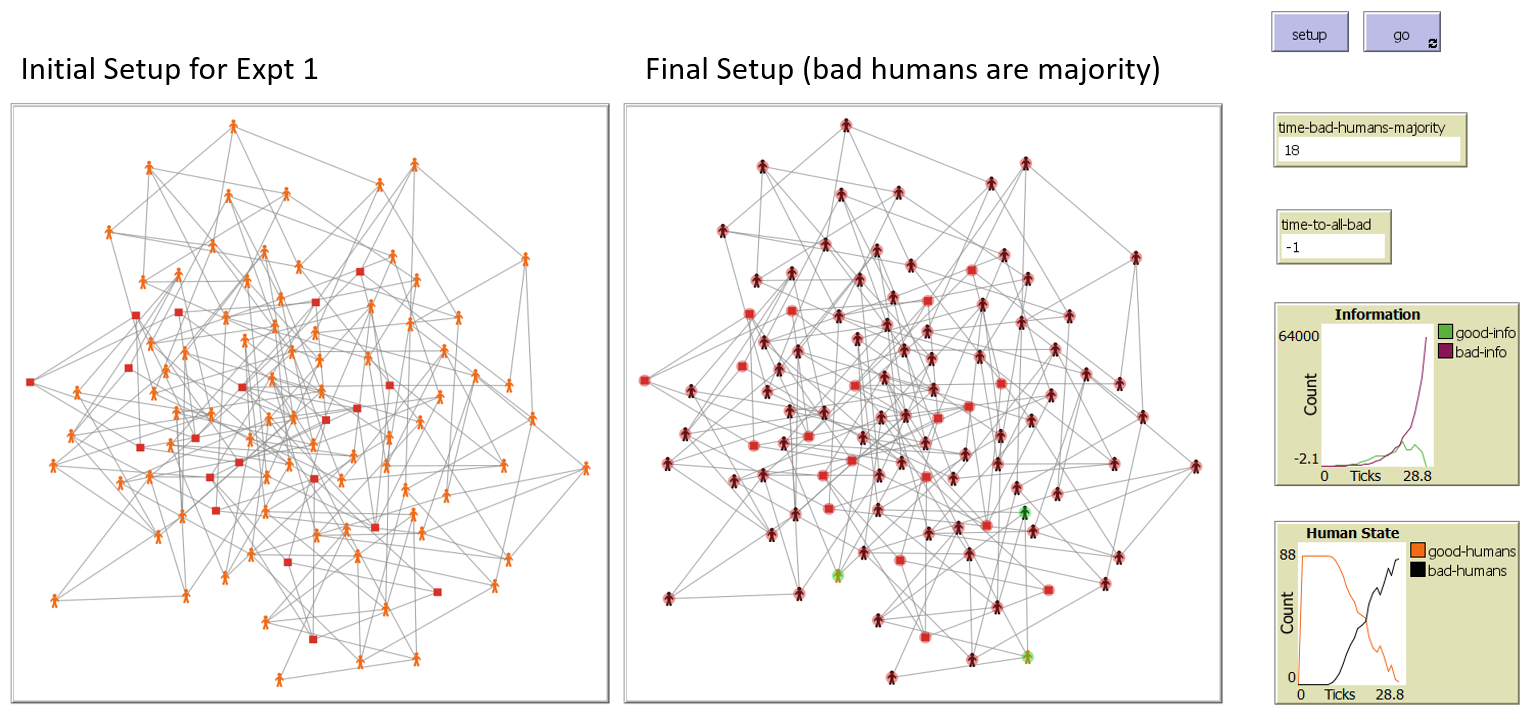} 
    \caption{Setup of the simulation model. In the initial setup, Humans are visualized with orange persons, and Bad Bots as red squares. In final setup, Bad Humans are colored black, and the green/red circles represent whether the current set of information consumption queue is good/bad.}
    \label{fig:setup}
\end{figure}

\begin{figure}[h!]
    \centering
    \includegraphics[width=1.0\textwidth]{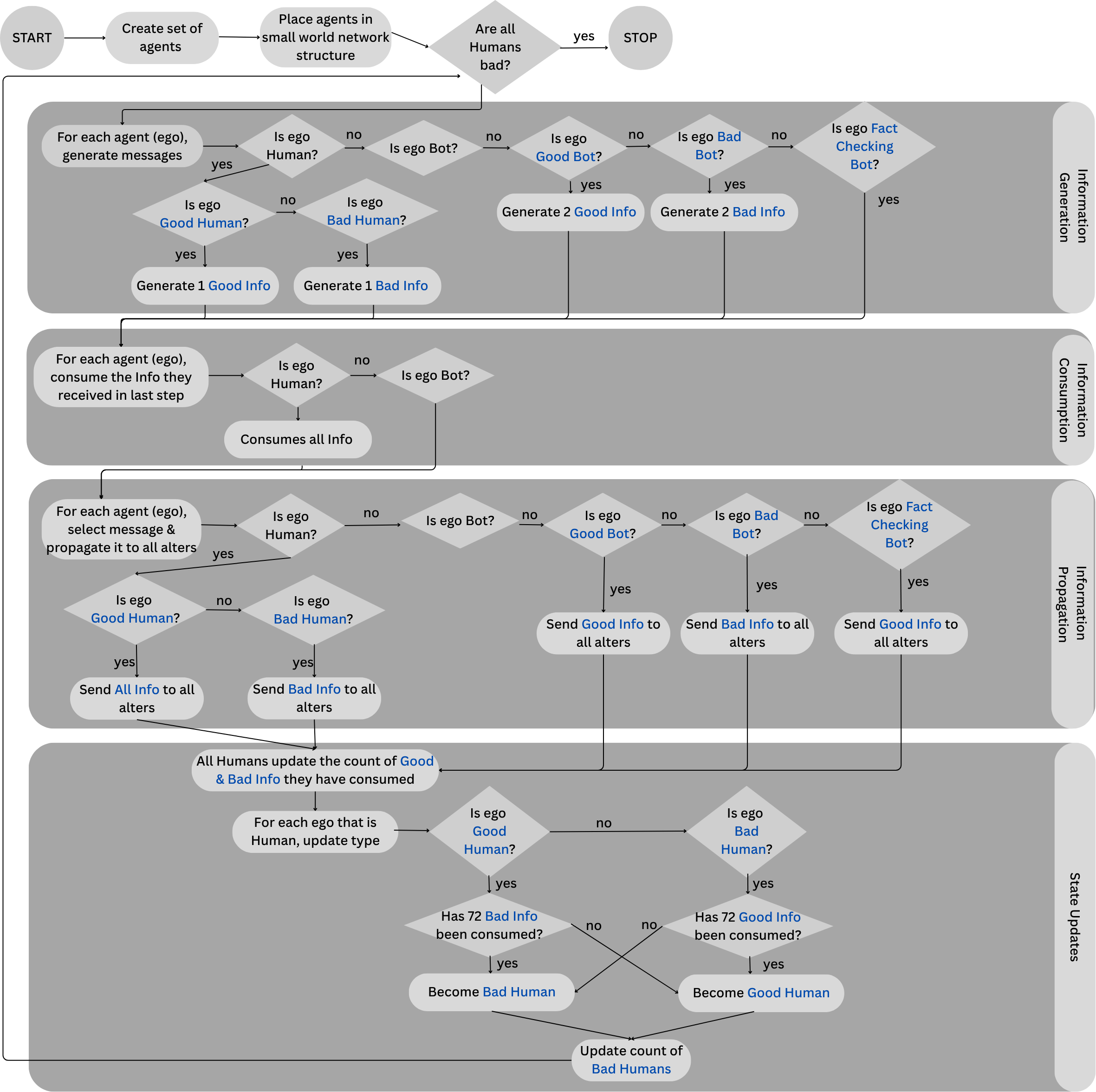} 
    \caption{Flowchart of Simulation Logic}
    \label{fig:flowchart}
\end{figure}

In our simulation, BotSim, we simulate the impact of information correction and the introduction of new good information in a social network. The model's main routine begins with creating a set of agents and placing them in a small world network structure. These agents consist of $n_h=1000$ Human agents and some of each type of Bots. The number of each type of Bot agents is a proportion of the number of Human agents, and is an independent variable varied during the experiments. Then, the simulation has four cyclical stages: Information Generation, Information Consumption, Information Propagation and State Update. These stages occur at each tick of the simulation until the simulation is stopped. Figure~\ref{fig:setup} contains screenshots of the network structure of the initial setup and the final setup when the simulation has stopped. Figure~\ref{fig:flowchart} illustrates the flowchart of simulation logic. The simulation is implemented in NetLogo.

The following steps summarizes the main routine. 

\begin{enumerate}
    \item The model is set up with a Small World network. The network is initialized with $n_h=1000$ Human agents and a number of Bad Bot agents $= \alpha_1\times$(Humans). This network is static over time, i.e., connections do not change over time. Links between agents $A_1$ and $A_2$ means that $A_1$ can send and receive information with $A_2$ and vice versa.
    \item The routine then enters a cycle. This cycle consists of three stages that repeat until a termination condition is met.
    
    \begin{enumerate}
        \item The Information Generation stage generates two types of information: Good Info and Bad Info. All agents generate Info (information) with a probability of $P_g$. Whether the agent generates Good or Bad Info depends on the state (i.e., Good, Bad) of the agent. Bots generate 2 pieces of Info while Humans generate 1 piece of Info.

        \item Next, we have the Information Consumption Stage, where agents consume the Info that they received in the previous tick, each with a probability of $P_c$. A Human consumes all Info, while Bots just hold the Info in their memory space. The consumption of Info represents the readership and processing of the information. Bots do not consume Info because they are pre-programmed with their state, and therefore are not influenced by incoming information.

        \item The third stage is the Information Propagation stage. Each agent (the ego) sieves through the messages they received in the previous tick to decide which to pass on to the alters in their ego network. Alters are neighboring agents attached to the ego through a direct link in the network. If the ego is a Good Human, all Info are selected for propagation. If the ego is a Bad Human, only Bad Info are selected. If the ego is a Good Bot or Info-Correction Bot, only Good Info are selected.  The ego then propagates each piece to its alters with a probability of $P_p$.
        
        \item The Status Update stage manages state transitions for Humans. These transitions happen when the amount of information the Human agent consumed hits a threshold. The threshold of Human agent state change is currently set to $t=72$ which is the average number of exposures to a piece of information a human need to form an opinion (i.e., belief). Bot agents are static (i.e., programmed with their belief) and do not experience state changes.
        
        \item The final stage in this cycle checks the conditions of the network, which tracks information consumption metrics by doing a count of good/bad information consumed per Human agent, and monitors the outcome measures to determine if the simulation should be terminated. The simulation terminates either when all Human agents turn into Bad Humans, or is stopped at 100 time ticks.   
    \end{enumerate}
\end{enumerate}

\subsection{Outcome Measures}
We measure two outcomes in this model: (1) \texttt{Bad Humans Majority}, which is the time required for the number of Bad Humans to be greater than the number of Good Humans; and (2) \texttt{All Bad Humans}, which is the time when all Humans are converted to Bad Humans.

The simulation's stopping criterion is when \texttt{All Bad Humans} is a valid positive value. This positive value measures the number of ticks where all Human agents are Bad Humans. This is akin to the formation of a conspiratorial society, where all humans spread conspiracies. If the simulation does not converge, we stop the simulation at 100 ticks. We set this simulation at 100 ticks based on preliminary sensitivity analysis and observations of the trend lines of the agent states' count. In these preliminary analyses, most of the simulated experiments either converged to a stable state or demonstrated clear directional trends by 100 ticks. Therefore, we use the 100 tick mark as a reasonable proxy for long-term outcomes, to balance computational efficiency with sufficient observation time to capture the emergent population behavior. A higher \texttt{Bad Humans Majority} measure is more desired, because it means the society takes a longer time for Bad Humans to dominate the conversation.

\subsection{Model parameters}
\label{app:keyparams}
Table~\ref{tab:key_parameters} reflects the key parameters used in our simulation model. These values were chosen on the basis of a combination of theoretical considerations and practical limitations. The model implementation relied on some stylized facts extracted from real-world data to provide theoretical validity to our model. These stylized facts and their implementation in our model are listed in \autoref{tab:stylized_facts_implementation}.

\begin{table}[h!]
    \centering
    \begin{tabular}{|p{2.5cm}|p{7cm}|p{2.5cm}|p{2.5cm}|}
    \hline
        \textbf{Parameter} & \textbf{Definition} & \textbf{Default Value} & \textbf{Range} \\ \hline
        $n_h$ & Number of Human agents & 100 & 100 \\ \hline 
        $p$ & Probability of formation of links between two agents in the starting network & 0.05 & 0.05 \\ \hline 
        $\alpha_1$ & Proportion of Humans that are Bad Bot agents & 0.2 & [0.1,1.0] \\ \hline 
        $\alpha_2$ & Proportion of Bad Bots that are Info-Correction Agents & 0 & [0.1, 1.0] \\ \hline 
        $\alpha_3$ & Proportion of Bad Bots that are Good Bots & 0 & [0.1, 1.0] \\ \hline 
        $P_g$ & Probability of generating information & 0.4 & 0.4 \\ \hline
        $P_c$ & Probability of an information being consumed & 0.8 & 0.8 \\ \hline 
        $P_p$ & Probability of an information being propagated & 0.8 & 0.8 \\ \hline 
        $t$ & Threshold of Human state change & 72 & 72 \\ 
    \hline
    \end{tabular}
    \caption{Key Parameters}
    \label{tab:key_parameters}
\end{table}

Our implementation mechanics of the number of Info that bots post is based on large-scale year long empirical analyses on bot and human behavior. In data from the 2020 coronavirus pandemic, bots generated at least two times more posts than humans \citep{ng2025social}. In data from the 2020 US Presidential Elections, bots generated about 2.6-3 times more posts than human users~\citep{luceri2020down}. In our simulation, we use the lower bound value where bots generate at least twice more posts than humans, to evaluate the scenario where bots are less aggressive.

Further, conspiracy theories often have a kernel of truth that malicious bots distort \citep{keeley1999journal}. This is reflected in the mechanic where Bad Bots convert Good Info into Bad Info. Confirmation Bias makes people more likely to spread information that aligns with their own beliefs \citep{hart2009feeling}, thus Bad Humans are more likely to propagate bad Info. People share information based on desirable conclusions, such as garnering attention and inducing humor rather, than accuracy \citep{moravec2019fake}, inspiring the behavior where Good Humans propagate both Good and Bad Info. Such behavior stems from the motivated reasoning bias \citep{moravec2019fake,kunda1990case}. The same behavior is also based on observational studies where people are ignorant of the validity of the information \citep{buchanan2020people}.

Finally, we operationalized the flip threshold $t=72$. Empirical human-subjects studies reveal that about 4.94 to 138.59 exposures to information on social media (i.e., text, image, video) can cause them to form a belief towards a topic. \citep{asher2018opinion}. We align our flip threshold $t$ with the mid range of the estimate from the empirical study.

\begin{table}[h]
    \centering
    \begin{tabular}{|p{5.5cm}|p{5.5cm}|p{4cm}|}
    \hline
        \textbf{Implementation within Simulation} & \textbf{Stylized Fact from Literature} & \textbf{Reference} \\ \hline
        Small-world network for conspiracy theory spread on social media & Religious ideologies and conspiracies on Twitter surrounding the 2015 Boston bombing incident self-organizes into a small-world phenomenon & \cite{ch2015local} \\ \hline


        Link probability of a small-world network is $p=0.05$ & Literature on modeling information dissemination and of rumor propagation constructs and measures small world networks with $p=0.05$ & \citep{fu2012information} \newline \citep{zanette2002dynamics} \\ \hline
        
        Bots generate 2x more Info than Humans & Bots post at least 2x more tweets than humans \newline Bots shared at least 2.6-3x the number of tweets than human users. & \cite{ng2025social} \newline \cite{luceri2020down} \\
        \hline

        Consumption rules: consuming about 70 pieces of information to flip states & Opinion formation threshold estimates: Empirical survey work show that individuals require about 4.94-138.59 exposures towards an item (text, image, video) before they form a stable opinion on a topic. & \cite{asher2018opinion} \\ \hline 

        Asymmetric propagation rules: Good Humans propagate both good and bad Info, Bad Humans propagate only bad Info & Motivated reasoning \& confirmation bias: People with stronger conspiracy beliefs are more uniform views \newline People with weaker conspiracy beliefs share more diverse views & \cite{pennycook2021psychology} \newline \cite{guess2019less} \\ \hline

        Bad Bots convert good Info into bad Info & Malicious bots distort facts to form disinformation & \cite{giachanou2022online}  \\ \hline

        Bad Bots convert good Info into bad Info & Conspiracy theories incorporate factual elements to enhance plausibility & \cite{keeley1999journal} \\ \hline 

        Bad Humans are more likely to propagate bad Info & Confirmation Bias: People are more likely to spread information that aligns with their own beliefs & \cite{hart2009feeling} \\ \hline

        Good Humans share both good and bad Info. & Motivated Reasoning: People share information based on desirable conclusions rather than accuracy \newline People cannot tell the validity of conspiracies. & \cite{moravec2019fake} \newline \cite{buchanan2020people} \\ \hline

        Bad Humans generate and propagate bad Info only & Monological belief system: Belief in one conspiracy theory predicts belief in others & \cite{goertzel1994belief}  \\ \hline 
    \end{tabular}
    \caption{Stylized Facts used in the Implementation}
    \label{tab:stylized_facts_implementation}
\end{table}

\section{Agent Mechanics Per Timestep}
The agent activation routine is a step-based advancement. We use the term ``tick" to represent a time step in the simulation, where the three cyclical stages take place. After each tick, the simulation moves to the next round. 

Within each tick, each agent has an activation routine that reflects its persona. For Human agents, there are two types: Good Humans and Bad Humans. Good Humans represent the generic network user; Bad Humans represent users who deliberately spread disinformation like conspiracy theorists. Humans generate 1 piece of Info per tick. Good Humans generate Good Info, Bad Humans generate Bad Info. Humans consume all Info. Good Humans propagate both Good and Bad Info, while Bad Humans propagate only bad Info. Humans change their state from good to bad or bad to good when the cumulative amount of Info they consume crosses a set threshold $t$. For all experiments, we use $t=72$, which is the average number of exposures that individuals require to form a stable opinion about a topic\citep{asher2018opinion}.

In our model, the design choice where Good Humans propagate both good and bad information and Bad Humans propagate only bad information reflects documented asymmetries in information sharing behavior. This leans on research that individuals with stronger conspiracy beliefs demonstrate greater selectivity in information sharing behaviors \citep{pennycook2021psychology}, and those with weaker conspiracy beliefs are more likely to share diverse content \citep{guess2019less}.

For Bot agents, there are three types: Bad Bots, Info-Correction Bots, and Good Bots. Bad Bots generate Bad Info, Good Bots generate Good Info. Bots generate Info at twice the rate of Humans. Bad Bots mimic malicious users that generate conspiracy theories and selectively propagate only conspiracy theories. Info-Correction Bots mimic the reactions of fact-checking organizations when they come across Bad Info: change them into Good Info (i.e., fact-check) and putting out only Good Info \citep{chang2021social}. Good Bots mimic the mechanics where organizations put out good messaging or advertisements instead of directly engaging with the malicious messages \citep{van2017inoculating}. Collectively, Info-Correction and Good Bots are useful bots.


\section{Virtual Experiments}
We ran five different virtual experiments, each designed to systematically explore different aspects of the effectiveness of useful bot interventions. The experiments can be grouped into two groups: single-parameter variation studies (Experiments 1 to 3) and dual-parameter sweeps (Experiments 4-5). In all the runs, we kept the probabilities used in information management ($P_c,P_g,P_p$), the number of Human agents ($n_h=1000$), and the Human state-change threshold ($t=72$) remains the same. This ensures that the observed differences in outcomes can be attributed to the different bot deployment strategies. All experiments were performed with Netlogo's BehaviorSpace function. Each parameter combination is performed $n=15$ times, and the mean time of the $n=15$ rounds is reported. We ran a total of 3,675 simulation runs across all experiments. Table~\ref{tab:virtual_experiments} summarizes the virtual experiments.

Experiment 1 serves as a baseline experiment, systematically varying only the proportion of Bad Bots ($\alpha_1$) from 0.1 to 1.0 in increments of 0.1, while maintaining zero useful bots in the system ($\alpha_2=\alpha_3=0$). This is represented in our Virtual Experiment table as [0.1, 1.0, 0.1]. That means that we tested the scenarios where the number of Bad Bots are 10\% of the number of Human agents through the scenario where the number of Bad Bots are 100\% the number of Human agents, each time increasing the percentage of Bad Bots by another 10\%. This experiment enables us to establish the fundamental relationship between malicious bot presence and the conspiratorial society formation in the absence of any countermeasures from useful bots. Experiment 2 introduces Info-Correction Bots as the sole intervention mechanism, while maintaining a fixed Bad Bot proportion ($\alpha_1=0.2$). This experiment varies the Info-Correction Bot ratio ($\alpha_2$) from 0.1 to 1.5 in increments of 0.1. The parameter range extends to $\alpha_2=1.5$ to explore the scenario where the Info-Correction Bots substantially outnumber Bad Bots. Similarly, Experiment 3 tests Good Bots in isolation by maintaining the Bad Bot proportion ($\alpha_1=0.2$) while varying the Good Bot ratio ($\alpha_3$) from 0.1 to 2.0 in increments of 0.1. The extended range to $\alpha_3=2.0$ tests whether proactive good messaging requires different optimal deployment ratios compared to reactive information correction. The design for Experiments 2 and 3 isolates the effectiveness of information correction and good messaging mechanisms respectively without confounding effects from other intervention types.

Experiments 4 and 5 uses dual-parameter sweeps to map interaction effects between Bad Bots and each useful bot type. Experiment 4 explores the Bad Bot: Info-Correction Bot parameter space. It varies $\alpha_1$ and $\alpha_2$ from 0.1 to 1.0, creating 10*10=100 unique parameter combinations. Experiment 5 explores the Bad Bot: Good Bot parameter space, varying $\alpha_1$ from 0.1 to 2.0 and $\alpha_3$ from 0.1 to 1.0.
These two experiments use parameter sweeps to enable analysis of optimal deployment ratios of useful bots.

To determine the number of replications required per experimental condition, we conducted a preliminary power analysis using one-way ANOVA models in Python. For each dependent variable (\texttt{Bad Human Majority}, \texttt{All Bad Humans}), we fitted a three Ordinary Least Square models with the independent variables $\alpha_1$, $\alpha_2$, $\alpha_3$ (the Bad Bots, Info-Correction and Good Bots as a proportion of the number of Humans respectively). From the ANOVA results, we computed the effect size ($\eta^2$) and converted it to Cohen’s $f$, then used the \texttt{FTestAnovaPower} function from \texttt{statsmodels} in Python to estimate the required sample size per group for 80\% power at $p=0.05$ level. \autoref{tab:power} presents the power analysis for the three independent variables. On average, the power analysis indicated that only $n=2$ replications per condition were sufficient to detect the observed effects; however, to mitigate stochastic variability inherent in ABM simulations, we conservatively ran $n=15$ replications for each condition. In total, we ran 3,675 experiments.

\begin{table}[h!]
    \centering
    \begin{tabular}{|p{3cm}|p{2.2cm}|p{2.2cm}|p{2.2cm}|p{2.2cm}|p{2.2cm}|}
    \hline
        \textbf{Expt No./ Params} & \textbf{1} & \textbf{2} & \textbf{3} & \textbf{4} & \textbf{5} \\ \hline 
        \textbf{Description} & Vary Bad Bot Proportion & Vary Info-Correction Bot Proportion & Vary Good Bot Proportion & Parameter Sweep (Bad Bot: Info-Correction Bot) & Parameter Sweep (Bad Bot: Good Bot) \\ \hline 
        \multicolumn{6}{|l|}{\textbf{Control Variables}} \\ \hline
        
        $n_h$, Number of Human Agents & 1000 & 1000 & 1000 & 1000 & 1000 \\ \hline 
        $P_g$, Probability of an agent generating information & 0.4 & 0.4 & 0.4 & 0.4 & 0.4 \\ \hline 
        $P_c$, Probability of information being consumed & 0.8 & 0.8 & 0.8 & 0.8 & 0.8 \\ \hline 
        $P_p$, Probability of information being propagated & 0.8 & 0.8 & 0.8 & 0.8 & 0.8  \\ \hline 
        $t$, threshold of Human state change & 72 & 72 & 72 & 72 & 72 \\ \hline 
        
        \multicolumn{6}{|l|}{\textbf{Independent Variables}} \\ \hline
        \textbf{$\alpha_1$} \newline (Bad Bots as a proportion of Humans) \newline [min, max, step] & $[0.1,1.0,0.1]$ & 0.2 & 0.2 & $[0.1,1.0,0.1]$ & $[0.1,2.0,0.1]$ \\ \hline 
        
        \textbf{$\alpha_2$} \newline (Info-Correction Bots as a proportion of Humans) \newline [min, max, step] & 0 & $[0.1,1.5,0.1]$ & 0 & $[0.1,1.0,0.1]$ & 0 \\ \hline 
        
        \textbf{$\alpha_3$} \newline (Good Bots as a proportion of Humans) \newline [min, max, step] & 0 & 0 & $[0.1,2.0,0.1]$ & 0 & $[0.1,1.0,0.1]$ \\ \hline

        \textbf{\# Replications} & 15 & 15 & 15 & 15 & 15 \\ \hline 

        \textbf{\# Conditions} & 10 & 15 & 20 & 10*10 = 100 & 10*10 = 100 \\ \hline 

        \textbf{Total Runs} & 150 & 225 & 300 & 1500 & 1500 \\ \hline 
        
        \textbf{Total \# experiments} & \multicolumn{5}{|c|}{3675} \\ \hline 
        
        \multicolumn{6}{|l|}{\textbf{Dependent Variables (Results)}} \\ \hline
        \texttt{Bad Human Majority} (Mean ticks) & $13.51\pm1.07$ & $17.85\pm2.27$ & $18.52\pm4.86$ \newline DNC when $\alpha_3\geq1.6$ & $17.73\pm3.62$ & $17.55\pm3.32$ \\ \hline 
        \texttt{All Bad Humans} (Mean ticks) & $22.82\pm2.26$ & $24.73\pm1.18$ & $27.06\pm2.84$ & DNC & DNC \\ \hline 
    \end{tabular}
    \caption{Summary of Virtual Experiments. DNC means that the run does not converge and was terminated at 100 ticks.}

    \label{tab:virtual_experiments}
\end{table}

\begin{table}[h!]
    \centering
    \begin{tabular}{|p{3cm}|p{2cm}|p{2cm}|p{3cm}|}
    \hline
    \textbf{Dependent Variable} & \textbf{$\eta^2$} & \textbf{Cohen's $f$} & \textbf{Runs/ Condition for 80\% power, $\alpha=0.05$} \\ \hline
    \multicolumn{4}{|l|}{\textbf{Vary Bad Bot ($\alpha_1$) Proportion}} \\ \hline
    \texttt{All Bad Humans} & 0.85 & 2.39 & 1.96 \\ \hline
    \texttt{Bad Human Majority} & 0.83 & 2.22 & 2.09 \\ \hline 
    \multicolumn{4}{|l|}{\textbf{Vary Info-Correction Bot ($\alpha_2$) Proportion}} \\ \hline
    \texttt{All Bad Humans} & 0.89 & 2.85 & 1.74 \\ \hline
    \texttt{Bad Human Majority} & 0.87 & 2.57 & 1.86 \\ \hline 
    \multicolumn{4}{|l|}{\textbf{Vary Good Bot ($\alpha_3$) Proportion}} \\ \hline
    \texttt{All Bad Humans} & 0.84 & 2.28 & 2.02 \\ \hline
    \texttt{Bad Human Majority} & 0.81 & 2.06 & 2.21 \\ \hline 
    \end{tabular}
    \caption{Power Analysis. Statistically, just two replications per condition would give 80\% power to the experiment at $p=0.05$.}
    \label{tab:power}
\end{table}

\section{Results}
\subsection{Varying one bot type}
\begin{figure}[h]
    \centering
    \includegraphics[width=\textwidth]{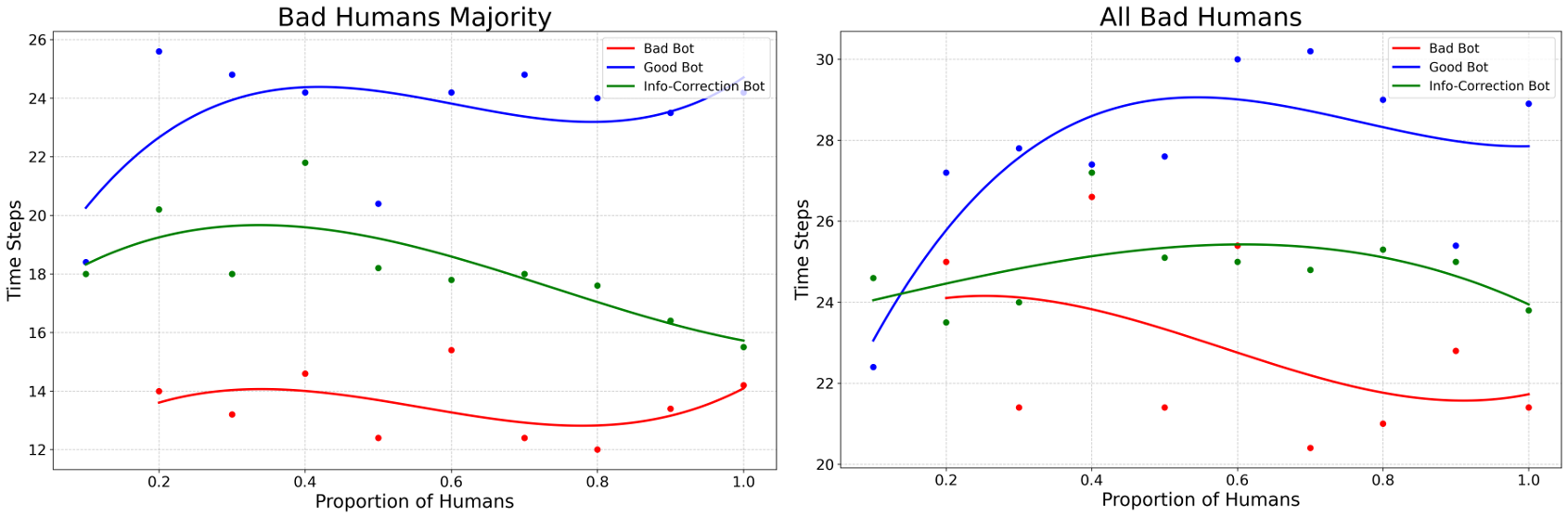} 
    \caption{Mean time to \texttt{Bad Humans Majority} and \texttt{All Bad Humans} from singularly varying the proportion of Bad Bots, Info-Correction Bots and Good Bots.}
    \label{fig:single_variation}
\end{figure}

Experiments 1, 2 and 3 vary only one bot type (Bad Bot, Info-Correction Bot, Good Bot), enabling isolation of individual bot effects. Figure~\ref{fig:single_variation} shows the mean time to \texttt{Bad Human Majority} in these three cases. This figure reveals distinct intervention effectiveness patterns: a stable baseline with rapid conspiracy formation (Expt 1), dramatically improved resistance with Info-Correction bot interventions (Expt 2), and the emergence of threshold effects where sufficient Good Bot deployment can prevent a conspiratorial majority entirely. The curves plotted in Figure~\ref{fig:single_variation} are not entirely smooth and straight because there is still a level of stochasticity in our model, i.e. the probability of generation, consumption and propagation of information, attachment to other agents during network formation. This stochasticity reflects the real-life probability distribution of the information interaction dynamics, rather than a constant interaction factor.

Experiment 1 establishes the baseline threat, where Bad Bots consistently drive conspiratorial society formation regardless of their proportion. This observation confirms that, when there are only Bad Bots in the society, without countermeasures, conspiratorial dominance is inevitable (mean time to \texttt{Bad Human Majority}: $13.51\pm1.07$ ticks). The society will eventually converge to a state where all Humans are bad (mean time to \texttt{All Bad Humans}: 22.82$\pm$2.26 ticks). Even small proportions of Bad Bots can reliably transform the information environment over time.

Both the injection of Info-Correction Bots (Experiment 2) and Good Bots (Experiment 3) can improve resistance to conspiracy formation, which can be seen from the graphs where more time steps are required for \texttt{Bad Human Majority} with the introduction of Good Bots ($18.52\pm4.86$ ticks) or Info-Correction Bots ($17.85\pm2.27$ ticks), as compared to when there were only Bad Bots ($13.51\pm1.07$ ticks). While the introduction of useful bots delay the formation of a conspiratorial society, a conspiratorial society still forms after a sufficiently long period of time. Good Bots demonstrate an additional capability over Info-Correction Bots. While Info-Correction Bots will always reach the \texttt{All Bad Humans} state, when $\alpha_3\geq1.6$, Good Bots prevent Bad Humans from reaching majority status. This suggests that proactive messaging may be more robust than reactive corrective strategies.

\subsection{Varying two bot types}
Experiments 4 and 5 are parameter sweep runs that explore the behavior space between the parameters of Bad Bots vs. Info-Correction and Bad Bots vs. Good Bots. These two experiments compare the effectiveness of using Info-Correction and Good Bots in combating conspiracy theories.
The results from these experiments tell us the optimal ratios between malicious and useful bots that maximize the time required for Bad Humans to dominate the conversation.

\begin{figure}[h!]
    \centering
    \includegraphics[width=1.0\textwidth]{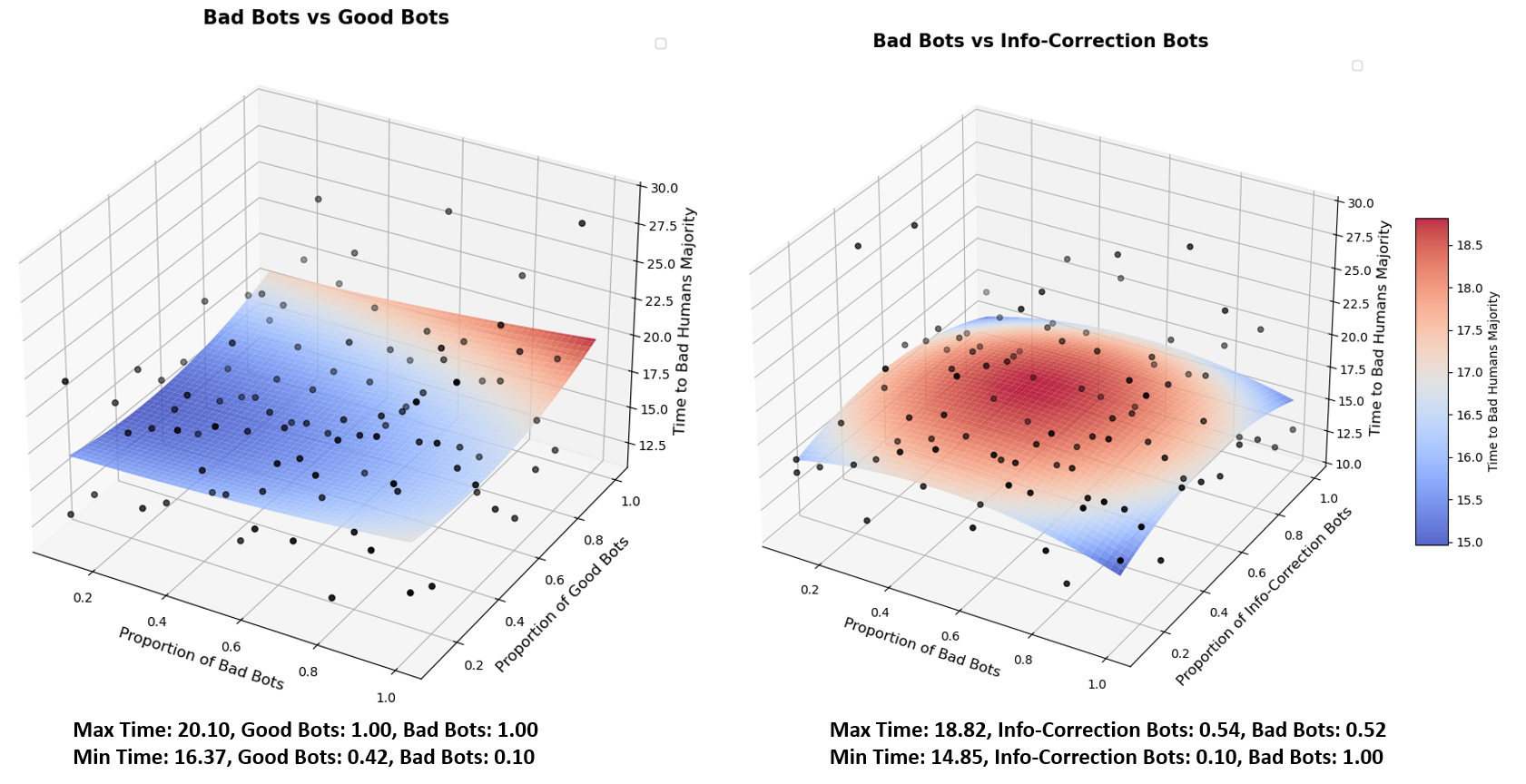} 
    \caption{Response Surface Analysis for varying Bad Bots and Good Bots vs Bad Bots and Info-Correction Bots. The $z$-axis represents time to \texttt{Bad Human Majority} in simulation ticks. The Info-Correction surface exhibits strong concavity ($2\beta_5=-18.6$), indicating diminishing returns with more Info-Correction Bots. The Good Bot surface is almost linear ($2\beta_5=+0.11$), indicating increasing benefits with increased number of Good Bots.}
    \label{fig:response_curves}
\end{figure}

We present the parameter sweep results as 3D response surface analysis in Figure~\ref{fig:response_curves}. We use a quadratic function as a smoothing function. While the underlying data points are discrete, the surface interpolation reveals continuous contours of influence across the behavior space, like the regions of peak effectiveness of each type of bot and surface topologies,  which provides insights into optimal useful bot deployment strategies.

For both surfaces, we calculated the defender efficiency functions as a quadratic response surface, to capture how the variations in defender  proportions (i.e., proportion of Good Bot, proportion of Info-Correction Bot) influence the time required for a \texttt{Bad Human Majority} scenario to emerge. These equations with their estimated coefficients are expressed in \autoref{eq:defender-efficiency}.

\begin{figure}[h]
\centering
\begin{align}
\text{Good Bots:} \quad 
T(b,d) &= 17.222 - 38.906b + 0.410d - 10.406bd + 503.362b^{2} + 0.0538d^{2}, \\
\text{Info-Correction Bots:} \quad 
T(b,d) &= 14.459 + 7.511b + 9.060d + 1.671bd - 8.098b^{2} - 9.313d^{2}.
\end{align}
\caption{Fitted defender efficiency functions for good bots and info-correction bots. 
Here, \(T(b,d)\) represents the time to bad-human majority, \(b\) is the proportion of bad bots, and \(d\) is the proportion of defender bots (either good or info-correction). 
Coefficients are estimated from quadratic response surface regressions of data from Experiments 4 and 5.}
\label{eq:defender-efficiency}
\end{figure}

The Info-Correction surface exhibits strong concavity ($2\beta_5=-18.6$), and so forms a dome that peaks near moderate densities of Info-Correction bots. This means that there is diminishing returns to the information environment once the Info-Correction Bots saturate the system. In contrast, the Good Bots surface has a mild convexity that is close to linear ($2\beta_5=+0.11$). This implies that Good Bot interventions scale predictably: as the number of Good Bots increase, the time delay to \texttt{Max Human Majority} increases.

Across both surfaces, the maximum and minimum points further illustrate the defender efficiency contrast. For Info-Correction Bots, the fitted surface peaks at approximately $T=18.82$ (Bad Bots$=0.52$, Info-Correction Bots$=0.54$), and falls to a minimum of $T=14.85$ (Bad Bots$= 1.00$, Info-Correction Bots$=0.10$), indicating that Info-Correction bots are most effective only at moderate densities ($\alpha_2=0.54$) and lose influence as the system becomes saturated with Info-Correction Bots. In contrast, the Good Bot surface reaches a higher maximum of $T=20.08$ (Bad Bots$= 1.00$, Good Bots$=1.00$). The surface also has a higher minimum of $T=16.37$ (Bad Bots$=0.10$, Good Bots$=0.42$). The elevated minimum and maximum suggest that Good Bot defenses are more consistently effective across the range of agent proportions. 

The response surface curves show that Good Bots are generally better than Info-Correction Bots. Good Bot interventions scale with the number of Good Bots employed, while Info-Correction Bots will hit a limit, after which there is diminishing returns. This phenomenon could be because Good Bots do not have the overhead of information verification. 
Finally, neither of the two experiments converges, i.e. reach a \texttt{All Bad Humans} state, indicating that the presence of useful bots prevents a conspiratorial society.

\subsection{Statistical Analysis}
With the results of varying one bot type, we also performed a two-way ANOVA to test the interaction of bot type and proportion of bot type. Bot type is represented as a discrete variable of three factors: Good Bot, Bad Bot, Info-Correction Bot. Proportion is described as a continuous variable. The ANOVA equation is presented in \autoref{eq:anova}, and one equation was ran for each outcome measure (\texttt{All Bad Humans} and \texttt{Bad Human Majority}).

\begin{figure}[h]
\centering
\begin{equation}
    \hat{Y}_{\text{outcome measure}} \sim C(\text{Bot Type}_i) * \alpha_i
\end{equation}
\caption{Two-way ANOVA to test the interaction between bot type and proportion. Outcome measure refers to \texttt{All Bad Humans} and \texttt{Bad Human Majority}. $\alpha_i$ is the number of bots of type$_i$ as a proportion of the number of humans.}
\label{eq:anova}
\end{figure}

The results of the variance analysis using ANOVA tests are presented in Table~\ref{tab:anova}. For both outcome measures (\texttt{All Bad Humans}, \texttt{Bad Human Majority}), there is a strong main effect of bot type ($F=16.03, p<0.001$ and $F=80.49, p<0.001$ respectively). This indicates that the different types of bots used can affect how quickly the simulation converges towards a conspiratorial society state. In contrast, the main effect of bot proportion was not significant (at the $p<0.05$ level) in both models, suggesting that simply increasing the number of bots without regard to their type does not reliably increase convergence time. There is marginal interaction between bot type and its proportion ($F=2.91, p=0.075$ for \texttt{All Bad Humans}; $F=2.72, p=0.087$ for \texttt{Bad Human Majority}), indicating that the influence of bot quantity may differ slightly depending on the role that the bots play within the information ecosystem~\citep{keller2019social,deb2018social}.

\begin{table}[h]
    \centering
    \begin{tabular}{|c|c|c|c|c|}
    \hline
    \multicolumn{5}{|l|}{\textbf{Dependent Variable: \texttt{All Bad Humans}}} \\ \hline 
    ~ & \textbf{sum squared }& \textbf{degree of freedom} & \textbf{F} & \textbf{Pr($>$F)} \\ \hline 
    C(bot type) & 109.79 & 2.0 & 16.02 & 4.40E-5*** \\ \hline 
    C(proportion) & 0.46 & 1.0 & 1.36E-1 & 7.16E-1 \\ \hline 
    C(bot type):C(proportion) & 19.83 & 2.0 & 2.91 & 7.49E-1 \\ \hline
    \multicolumn{5}{|l|}{} \\ \hline 
    \multicolumn{5}{|l|}{\textbf{Dependent Variable: \texttt{Bad Human Majority}}} \\ \hline 
    ~ & \textbf{sum squared }& \textbf{degree of freedom} & \textbf{F} & \textbf{Pr($>$F)} \\ \hline 
    C(bot type) & 460.92 & 2.0 & 80.49 & 4.12E-11*** \\ \hline 
    C(proportion) & 1.10 & 1.0 & 3.78E-1 & 5.45E-1 \\ \hline 
    C(bot type):C(proportion) & 15.77 & 2.0 & 2.72 & 8.69E-2 \\ \hline
    \end{tabular}
    \caption{Two- way ANOVA analysis of proportion of each Bot Type and interaction of Bot Type and proportion for dependent variables. *** indicates significant at the $p<0.001$ level.}
    \label{tab:anova}
\end{table}

Next, with the overall results, we performed an Ordinary Least Squares (OLS) regression to understand the effect sizes of the proportion of each type of bot on the outcome measures. This regression formula is presented in \autoref{eq:overall_regression}. We ran one equation for each outcome measure (\texttt{All Bad Humans} and \texttt{Bad Human Majority}).

\begin{figure}[h]
\centering
\begin{equation}
\begin{aligned}
    \hat{Y}_{\text{Outcome Measure}} \sim\;
    & \beta_0 \\
    & + \beta_1 (\text{proportion of Bad Bots}) 
    + \beta_2 (\text{proportion of Good Bots}) \\
    & + \beta_3 (\text{proportion of Info-Correction Bots}) \\
    & + \beta_4 C(\text{Bad Bots Present})) 
    + \beta_5 C((\text{Good Bots Present})) 
    + \beta_6 C(\text{Info-Correction Bots Present}) \\
    & + \beta_7 (\text{proportion of Bad Bots} \times C(\text{Good Bots Present})) \\
    & + \beta_8 (\text{proportion of Bad Bots} \times C(\text{Info-Correction Bots Present})) \\
    & + \beta_9 (\text{proportion of Bad Bots} \times \text{proportion of Good Bots}) \\
    & + \beta_{10} (\text{proportion of Bad Bots} \times \text{proportion of Info-Correction Bots}) + \varepsilon
\end{aligned}
\end{equation}
\caption{Ordinary Least Squares regression model for predicting the outcome measures based on the starting proportion of agents. Outcome measure refers to \texttt{All Bad Humans} and \texttt{Bad Human Majority}.}
\label{eq:overall_regression}
\end{figure}

The results of the OLS regression are presented in \autoref{tab:overall_regression}. The first model for the dependent variable \texttt{All Bad Humans} has $R^2=0.198$. While the model's overall explanatory power is modest, several interaction effects are significant. The presence of Bad Bots interacting with Good Bots ($\beta=20.47, p=0.045$) and Bad Bots interacting with Info-Correction Bots ($\beta=-8.09, p=0.003$) show that useful bot mechanisms can substantially influence the convergence time of a conspiratorial society. The second model for explaining \texttt{Bad Human Majority} has $R^2=0.511$, which shows considerably stronger explanatory power. Here, the interaction between Bad and Good Bots ($\beta=42.31,p<0.001$) emerges as the most influential predictor, suggesting that the presence of Good Bots will dramatically reduce the formation of a conspiratorial society. Together, the two models reveal a non-linear influence architecture of each type of the bot on the information ecosystem. Rather, it is the interaction of Bad Bots with useful bots that yield emergent effects on the formation, or intervention of, a conspiratorial society.

\begin{table}[h]
\centering
\renewcommand{\arraystretch}{1.1}
\setlength{\tabcolsep}{6pt}
\begin{tabular}{|l|c|c|c|}
\hline
\multicolumn{4}{|l|}{\textbf{Dependent Variable: \texttt{All Bad Humans}}} \\ \hline
\multicolumn{4}{|l|}{$R^2 = 0.198$ \quad Adjusted $R^2 = 0.134$ \quad $F(9,112)=3.08$, $p=0.002$} \\ \hline
~ & \textbf{Coef.} & \textbf{Std. Err.} & \textbf{P-value} \\ \hline
Intercept & 25.06 & 2.14 & 0.000*** \\ \hline
Good Bots Present & -1.30 & 2.74 & 0.637 \\ \hline
Info-Correction Bots Present & -1.64 & 2.39 & 0.495 \\ \hline
Proportion of Bad Bots & -3.73 & 3.27 & 0.256 \\ \hline
Proportion of Good Bots & -0.19 & 0.08 & 0.011** \\ \hline
Proportion of Info-Correction Bots & 4.30 & 1.64 & 0.010** \\ \hline
Bad Bots $\times$ Good Bots Present & 20.47 & 10.07 & 0.045** \\ \hline
Bad Bots $\times$ Info-Correction Bots Present & 6.73 & 3.71 & 0.073 \\ \hline
Bad Bots $\times$ Proportion of Good Bots & 0.07 & 0.05 & 0.205 \\ \hline
Bad Bots $\times$ Proportion of Info-Correction Bots & -8.09 & 2.70 & 0.003** \\ \hline
\multicolumn{4}{|l|}{} \\ \hline
\multicolumn{4}{|l|}{\textbf{Dependent Variable: \texttt{Bad Human Majority}}} \\ \hline
\multicolumn{4}{|l|}{$R^2 = 0.511$ \quad Adjusted $R^2 = 0.471$ \quad $F(9,112)=12.99$, $p=5.42E-14$} \\ \hline
~ & \textbf{Coef.} & \textbf{Std. Err.} & \textbf{P-value} \\ \hline
Intercept & 13.89 & 1.84 & 0.000*** \\ \hline
Good Bots Present & 0.44 & 2.36 & 0.854 \\ \hline
Info-Correction Bots Present & 1.62 & 2.06 & 0.433 \\ \hline
Proportion of Bad Bots & -0.63 & 2.82 & 0.823 \\ \hline
Proportion of Good Bots & -0.46 & 0.07 & 0.000*** \\ \hline
Proportion of Info-Correction Bots & 1.53 & 1.41 & 0.281 \\ \hline
Bad Bots $\times$ Good Bots Present & 42.31 & 8.68 & 0.000*** \\ \hline
Bad Bots $\times$ Info-Correction Bots Present & -0.25 & 3.20 & 0.937 \\ \hline
Bad Bots $\times$ Proportion of Good Bots & 0.03 & 0.05 & 0.480 \\ \hline
Bad Bots $\times$ Proportion of Info-Correction Bots & -1.42 & 2.33 & 0.545 \\ \hline
\end{tabular}
\caption{Coefficients of the Ordinary Least Squares regression models predicting convergence outcomes. The upper panel corresponds to \texttt{All Bad Humans}, and the lower panel to \texttt{Bad Humans Majority}. ** indicates significance at $p<0.05$ and *** at $p<0.001$. Interaction terms represent moderation effects between the proportions or presences of different bot types.}
\label{tab:overall_regression}
\end{table}

\subsection{Validation}
Our results are validated through stylized facts that are used to determine known phenomena, providing pattern and theoretical validity to our model. These stylized facts are adapted from social science and social psychology theories, and are presented in Table~\ref{tab:stylized_facts}. 

In our simulation, Bad Humans generate and propagate Bad Info only. This is based on the concept of the monological belief system, where the belief in one conspiracy theory predicts and predates the belief in other conspiracy theories \citep{goertzel1994belief}. The same stylized fact also reflects how, when a simulation reaches \texttt{Bad Human Majority} state, it does not fall back to a state where Bad Humans are the minority. The information propagation rules in our simulation are asymmetric: Good Humans propagate Good and Bad information, while Bad Humans propagate only Bad information. This is motivated by sociological theories where people with stronger conspiracy beliefs share more uniform views, and those with weaker conspiracy beliefs share more diverse views \cite{pennycook2021psychology,guess2019less}.

In a social network, the combination of the pressure of social influence and the formation of echo chambers maintains coherence among conspiracy theories, which makes turning Bad Humans back to Good Humans difficult. Finally, our model reaches the \texttt{Bad Human Majority} state regardless of the presence of useful bots because disinformation spreads faster than truth \citep{vosoughi2018spread}.

\begin{table}[h]
    \centering
    \begin{tabular}{|p{4.2cm}|p{5.5cm}|p{5.5cm}|}
    \hline
        \textbf{Reference} & \textbf{Stylized Fact} & \textbf{Simulation Result} \\ \hline
        \cite{goertzel1994belief} \newline \cite{tornberg2018echo} & Belief in one conspiracy theory predicts belief in others \newline Formation of echo chambers as a complex contagion & After the simulation reaches \texttt{Bad Human Majority} state, it does not fall back to a state where Bad Humans are minority \\ \hline
        \cite{vosoughi2018spread} & Disinformation spread faster than truth & Simulation reaches \texttt{Bad Human Majority} state regardless of presence of useful bots \\ \hline
        \cite{lewandowsky2021countering} & Proactive inoculation is significantly more effective than reactive corrections & Good Bots are more effective than Info-Correction Bots \\ \hline 
    \end{tabular}
    \caption{Stylized Facts used for Validation}
    \label{tab:stylized_facts}
\end{table}

\section{Discussion}
This work simulates the information distribution mechanics in a social network, demonstrating the effects of malicious Bad Bots on the state of Humans, and the usefulness of Info-Correction and Good Bots on preventing an entirely conspiratorial society (which is represented in the simulation as all Humans being Bad). Our simulation design distinguishes itself from previous models of information propagation \citep{gurung2025modeling,gomez2013modeling} in three ways. First, most existing simulations examining conspiracies \citep{blane2021simulating,gurung2025modeling} and countermeasures \citep{murdock2025simulating,addai2025modeling} do not consider the presence of bots. We explicitly model different types of bots with distinct behavioral rules, reflecting our claim that the social media space can contain both malicious and beneficial automated agents \citep{assenmacher2020demystifying,ng2024cyborgs}. Second, the state-change mechanism for Human agents is based on cumulative information exposure, capturing the gradual nature of the process of conspiracy belief formation \citep{prooijen2018psychology,van2018conspiracy}. Third, we incorporate asymmetric information propagation rules that reflect empirical findings about selective sharing behaviors, and emphasize the asymmetric reality of social media-based warfare \citep{tucker2018social}.

Experiment 1 shows that even a small ratio ($\alpha_1$=0.1) is sufficient for the formation of a conspiratorial society, where all humans are bad. This mirrors past work that shows that about 20\% of bots are required to push a society towards the tipping point where the majority of opinions align with the bots' opinions \citep{carragher2023simulation}. This small ratio reflects the danger of the presence of Bad Bots spreading harmful information. 

In all replications of Experiment 1, the simulation will eventually reach a state where Bad Humans are the majority. That means that if the number of Bad Bots is left unchecked, a society will always result in a larger number of Bad Humans than Good Humans. This stresses the importance of the regulation of malicious agents on social media. One way of combating malicious agents is through mass banning. Platforms like X have been continually taking down malicious agents \cite{xtakedown}. However, such large-scale take downs are limited to the social media companies, and with unknown frequencies. What can citizen-based, non-profits and government organizations do to reduce the impact of malicious messaging on the public?

\begin{figure}[h]
    \centering
    \includegraphics[width=0.6\textwidth]{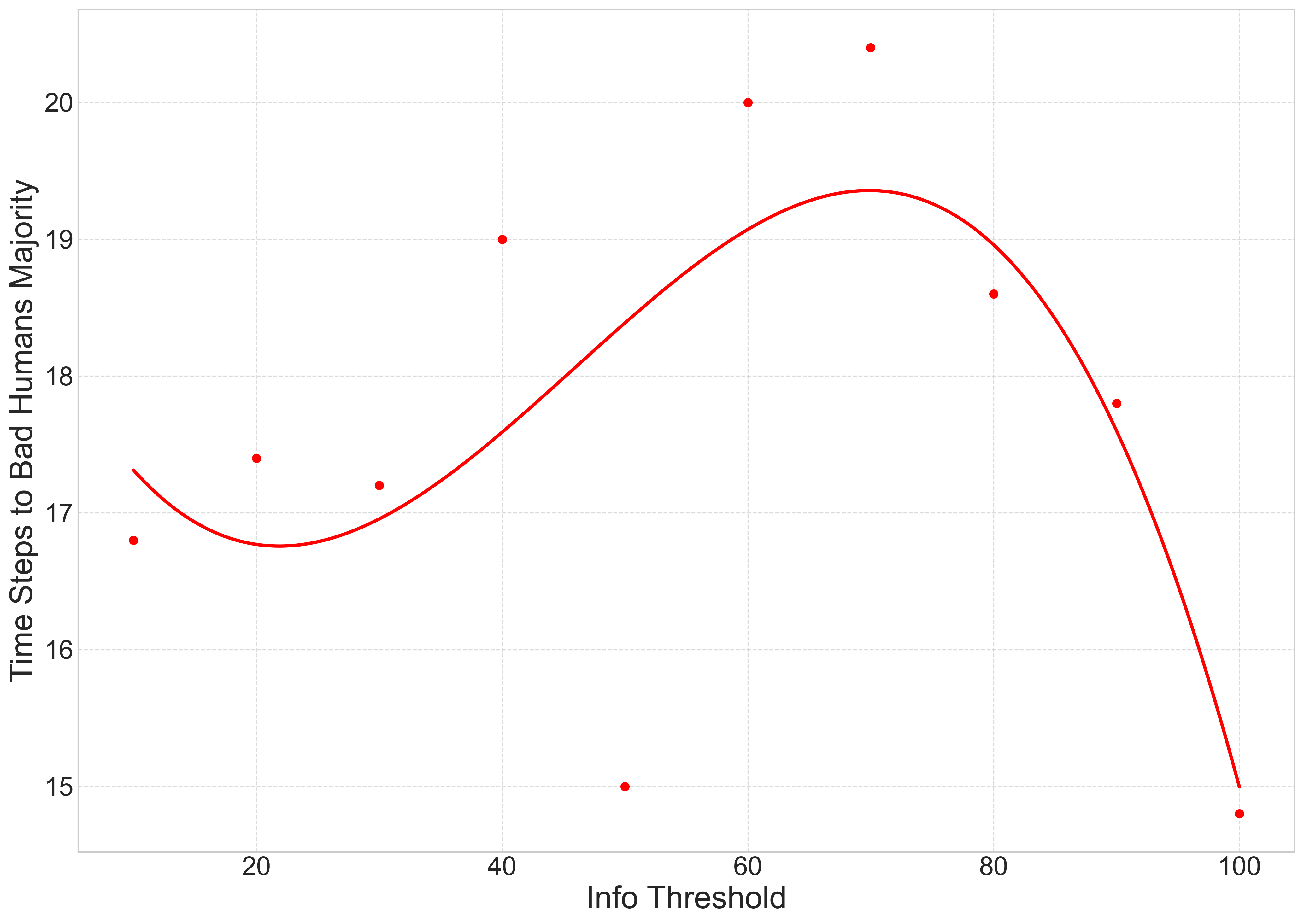} 
    \caption{Mean time to \texttt{Bad Humans Majority} from varying the Info Threshold from [10,100] with only Humans and Bad Bots ($n_h=1000$ and $\alpha_1=0.2,\alpha_2=0,\alpha_3=0$).}
    \label{fig:info_threshold}
\end{figure}

One common method that is being deployed is media literacy training, where governments and organizations try to teach users to spot Bad Info. The corollary of this intervention in our simulation is to vary the Info Threshold value. If Humans are more resilient to information injects, their Info Threshold value increases. Figure~\ref{fig:info_threshold} varies the Info Threshold from $t=10$ to $t=100$ with steps of $t=10$. Unfortunately, the relationship between Info Threshold and the time for \texttt{Bad Humans Majority} is non-linear. Initially, modest increases in threshold produce mid range values, but there eventually is a tipping point of $t=74$, beyond which, higher thresholds do not necessarily yield a more resilient society. In fact, overly resilient agents slow the overall information flow in the system. The inverted U-shape curve is also consistent with findings that overly skeptical audiences can disengage with corrective information~\citep{nyhan2019roles,lewandowsky2012misinformation}, and audiences that have already believed in conspiracy theories will be harder to be persuaded to change their beliefs.

In real life, not all humans can be trusted to achieve great information literacy, and from our experiments, changing the Info Threshold is not a guaranteed solution, so there must be a consideration for other types of interventions. Our subsequent experiments 2-5 demonstrate the feasibility of interventions using useful bots.

Experiments 2 and 3 show that the presence of useful bots (Info-Correction Bots and Good Bots) can combat harmful information by delaying or preventing the formation of conspiratorial societies (i.e., \texttt{All Bad Humans} state does not occur). In real life, activities similar to the mechanics of Info-Correction and Good Bots include: performing information corrective activities such as fact-checking or debunking disinformation, or carrying out educational campaigns like the \#WashYourHands campaign that ran during the 2020 coronavirus pandemic. These actions can be performed by organizations independent of social media companies, which provides such organizations more power to respond to bad information. Useful bots can prevent a conspiracy society, as observed by the lack of occurrence of \texttt{All Bad Humans} state. 

Statistical analyses through ANOVA and OLS both demonstrate that it is the type of bot rather than the proportion of bot that is the key determinant of convergence dynamics. The ANOVA statistical analysis (\autoref{tab:anova}) of the interaction between bot type and proportion suggest that the nature of the bot (i.e., Bad Bot, Good Bot, Info-Correction Bot) matters more than the absolute number of bots introduced. Further, the OLS regression analysis (\autoref{tab:overall_regression}) emphasize that bot types cannot be viewed in their singularity. It is their interactions with each other that influences the eventual trajectory of the network. These findings reinforce that the success of interventions on harmful information is not merely a function of the scale of the deployed bots but also the functional role of the agents and their network topology.

Experiments 4 and 5 compare the effect of using Info-Correction Bots against Good Bots. We compare the ratios of malicious to useful bots at the maximum time required for \texttt{Bad Humans Majority}. A longer time for Bad Humans to be the majority is more desired, to buy time for authorities to carry out educational campaigns or initiate laws (i.e., Singapore's POFMA that requires authors to addendum to their post, or the United States' FCC, which prohibits false information broadcasting). Our results show that the Good Bots exhibit a higher maximum time, and is therefore more desirable to Info-Correction Bots.

From the calculations of defender efficiency, Good Bots are more efficient than Info-Correction Bots. As the number of Good Bots increase, the time $T$ to \texttt{Bad Human Majority} increases. For Info-Correction Bots, as the number of Info-Correction Bots increases, the time $T$ increases up to a certain point, after which, the time $T$ decreases. This difference in efficiency arises from the different modes of action of the two types of useful Bots. Good Bots generate and propagate Good Info, which adds to the amount of Good Info in the information ecosystem, increasing the amount of Good Info that is likely to be recommended and read by Humans. Info-Correction Bots do not generate information, but instead when hit with Bad Info, will correct them to Good Info. They are therefore limited by the number of Info they receive to perform their tasks, and also consume more resources to perform the fact checking process.

Cognitive effort theory means that every action or decision makes requires a cognitive process and response \citep{westbrook2015cognitive}. In a bot, the cognitive effort can be measured by the programming steps required. From a cognitive effort point of view, Good Bots require less effort to construct as compared to Info-Correction Bots. The main action of Good Bots is the creation and dissemination of good information. If a Good Bot sends a message, it takes only three steps: decide that it is time to send a message, craft a message, then send the message. Info-Correction Bots require cognitive power to find mis/disinformation, analyze the information, fact-check the information, and create a coherent and convincing response against the misinformation. A fact-checking bot takes more programming steps for an Info-Correction Bot to complete the task, and its task involves more decision points which does consume resource, and higher chances where the task will default without completion. It is therefore easier and more predictable to automate the programming of Good Bots than Info-Correction Bots. In fact, currently, information correction tasks are still performed manually through fact-checking organizations and consortia, and it is a time-consuming and costly process \citep{lee2023fact}.

Rather than trying to stop or prevent Bad Bots from disseminating disinformation, evidence-based communication strategies can be used to slow the rate of the formation of a conspiratorial society. Communication strategies should leverage on useful bots (i.e., Info-Communication, Good Bots) to distribute good narratives en-masse. Further, good messaging is more resource effective than information correction, and can be more easily harnessed through automated information distribution with bots.  Currently, organizations and governments use information correction and good messaging techniques against bad information. Our simulation urges the use of automated bots to aid these efforts. Although different societal and cultural factors will require a unique strategy to address disinformation, a greater understanding of the process by which a conspiratorial society occurs and the role played by useful bots can help improve public strategies to address this societal concern. Appropriately designed engineering and architecture development of social media platforms can balance the human side of fact-based messaging with the use of these Artificial-Intelligent agents to correct information in real time.

\subsection{Limitations and Future Directions}
A limitation of our model is the assumption that agents consume and propagate information from their neighbors with an equal probability. In the real world, the probability differs among agents as a reflection of the strength of ties and biases present in agents. For example, homophily bias indicates that agents might consume more information that originates from other agents with similar demographic characteristics, and authority bias might result in information from authority figures being propagated with higher probability \citep{ng2024exploring}. 
Future work can combine the power of agent-based modeling with network science and calculate different probabilities for the information management stage based on agent properties and network ties. This includes studying the effects of the positions of useful agents, that is, whether placing useful bots strategically in the network mitigates bad information quickly.

Our model assumes that all information is about the same topic, and that the information propagated by useful bots is counter messages to the Bad Info propagated by Bad Bots. Further extensions of this work include studying the effects of timing in a pool of different topics, and how fast after a Bad Bot propagates Bad Info must useful bots respond to mitigate the conspiratorial effects.

Lastly, this project tested only two types of bots: Good Bots and Info-Correction Bots. In reality, there are a suite of social bots that could intervene the online space and possibly prevent conspiratorial society from forming~\citep{ng2025dual,gorwa2020unpacking}. Future research includes increasing the generalizability and robustness of our conclusions through testing a wider range of agent types.

\section{Conclusion}
Social media bots can be used as vessels of disinformation spread to induce conspiratorial societies, but more importantly, they can be used as mechanisms to prevent the formation of such societies. We construct and present BotSim, which simulates the interaction between Bad Bots that spread bad Info, and useful bots that aim to counter the effects of disinformation spread. Useful bots in the form of Info-Correction Bots turn bad Info to good, correcting the conspiracy theory, while Good Bots simply disseminate good Info.

Overall, our simulation experiments show that: 1) Bad Bots alone form a conspiratorial society; 2) Good Bots are 7\% more resource-effective than Info-Correction Bots for preventing the spread of conspiracy theories; and 3) the type of useful bot and its interactions with Bad Bots are key factors in determining their usefulness as interventions for harmful information.

This research offers an insight into the interplay between malicious bots and the use of bots. It sets the premise to leverage social media bots for information correction and good messaging. It further shows that good messaging is more resource effective than information correction, and can be more easily harnessed through automated information distribution with bots. We hope our work will increase the appreciation of social media bot capabilities, and spur discussions within public organizations towards integrating bot techniques in their strategic communications strategies. 

We make two broad contributions. Theoretically, we expand the concept of social media bots from a malicious agent to an automated agent that can be used for both social good and bad. This ties into the concept of a Cyber Social Agent, which encompasses the set of agents that can be used for both good and bad, and especially that the good can come from appropriate use of the same mechanics \citep{ng2025dual}. Useful agents, such as Information Correction Bots and Good Bots, can be used for information de-bunking and pre-bunking respectively, which eventually prevents a conspiratorial society. Empirically, we demonstrate this concept through the strength of an agent-based model to study how individual agents (humans and/or bots) interact within a network to potentially impact the behavior of larger populations (i.e., the formation of a conspiratorial society). Our BotSim model establishes a foundation for future work for characterizing the impact of bots in the online societal space.

\section{Acknowledgments}
This material is based upon work supported by the Scalable Technologies for Social Cybersecurity, U.S. Army (W911NF20D0002), Office of Naval Research (N000142112765), Threat Assessment Techniques for Digital Data, Office of Naval Research (N000142412414), and Community Assessment, Office of Naval Research (N000142412414). The views and conclusions contained in this document are those of the authors and should not be interpreted as representing official policies, either expressed or implied by the Office of Naval Research, U.S. Army or the U.S. government.

\endparano



% ...





\bibliographystyle{jasss}
\bibliography{references}


\end{document}